\def\year{2022}\relax
\documentclass[letterpaper]{article} % DO NOT CHANGE THIS
\usepackage{aaai22}  % DO NOT CHANGE THIS
\usepackage{times}  % DO NOT CHANGE THIS
\usepackage{helvet}  % DO NOT CHANGE THIS
\usepackage{courier}  % DO NOT CHANGE THIS
\usepackage[hyphens]{url}  % DO NOT CHANGE THIS
\usepackage{graphicx} % DO NOT CHANGE THIS
\usepackage{natbib}  % DO NOT CHANGE THIS AND DO NOT ADD ANY OPTIONS TO IT
\usepackage{caption} % DO NOT CHANGE THIS AND DO NOT ADD ANY OPTIONS TO IT
\DeclareCaptionStyle{ruled}{labelfont=normalfont,labelsep=colon,strut=off} % DO NOT CHANGE THIS
\usepackage{algorithm}
\usepackage{algorithmic}
\usepackage{amsmath}

\usepackage{subcaption}
\usepackage{newfloat}
\usepackage{listings}
\floatstyle{ruled}
\newfloat{listing}{tb}{lst}{}
\floatname{listing}{Listing}
\usepackage{mycomments3}
\title{Consistency Based Weakly Self-Supervised Learning for Human Activity Recognition with Wearables}
\author {
    Taoran Sheng and
    Manfred Huber
}
\affiliations {
    Department of Computer Science and Engineering\\
    University of Texas at Arlington\\
    taoran.sheng@mavs.uta.edu, huber@cse.uta.edu
}

\usepackage{bibentry}
\begin{document}

\maketitle

\com{We should probably empty out the author information rather than leaving the default. }

\begin{abstract}
While the widely \del{equipped}\add{available} embedded sensors in smartphones and other wearable devices make it easier to obtain data of human activities, recognizing different types of human activities from sensor based data, \del{on the other hand,} remains a difficult research topic in ubiquitous computing. One reason for this is most of the collected data is unlabeled. However, many current human activity recognition (HAR) systems are based on supervised methods, which heavily rely on the labels of the data . We describe a weakly self-supervised approach in this paper that consists of two stages: (1) In stage one, the model learns from the nature of human activities by projecting the data into an embedding space where similar activities are grouped together; (2) In stage two, the model is fine-tuned using similarity information in a few-shot learning fashion using the similarity information of the data. This allows downstream classification or clustering tasks to benefit from the embeddings. Experiments on three benchmark datasets demonstrate the framework's effectiveness and show that our approach can help the clustering algorithm achieve comparable performance in identifying and categorizing the underlying human activities as pure supervised techniques applied directly to \del{the} \add{a corresponding fully labeled}\del{original} \com{you did not talk about the way you got the data you used and thus it might not be clear what "original" here means.} data set.
\end{abstract}

\section{Introduction}
\noindent Many machine learning algorithms have been studied to address a variety of problems associated with wearable sensor-based human activity analysis. While the majority of them used supervised algorithms, largely relying on labeled data, a limited number of works have explored unsupervised or semi-supervised ways to address\del{ing} the recognition problem. However, since labeled data is in short supply and the bulk of acquired data is unlabeled, utilizing supervised techniques to recognize distinct human behaviors remains a difficult research challenge in ubiquitous computing. On the other hand, unsupervised approaches do not need labeled data to train the model; nevertheless, their performance is often inferior than that of supervised methods.

To overcome the aforementioned challenges, we aim at developing an approach that makes use of domain knowledge of human behavior, a limited amount of information about the similarity of data samples, and a \com{I am not sure if massive is the right word to use here. Maybe large or significant would be better. Massive suggests a very very large number.} \add{significant}\del{massive} quantity of unlabeled data. The proposed approach is motivated by two factors: first, the use of domain knowledge and a large amount of unlabeled data to simplify the model's learning task and force the model to retain useful task-specific features; \del{and} second, limiting the supervision required to train the model by relying \madd{only} on the similarity of a small amount of data samples rather than explicitly using the labels. Experiments have been conducted to assess the efficacy of our proposed approach.

\section{Related Work}
\subsection{Human Activity Recognition}\com{This section is somewhat short and it would be great to have some newer references (your newest are from 2016)}
There are two primary directions in wearable sensor-based \add{human activity recognition (}HAR\add{)}: approaches based on handcrafted features, and approaches based on deep neural networks \add{(DNN)}. In the former one, handcrafted features are designed using domain knowledge. For instance, \cite{sbhar,transitionAware} included statistical features such as mean, variance, and entropy into their models. In \cite{waveletFeature} , features taken from a wavelet transform were used. He and Jin \cite{dctFeature} extracted features using the discrete cosine transform. These features have the benefit of being readily derived from the signal and have been shown to be pretty successful in the HAR system. 

Various HAR models have recently used DNN to enable automatic feature extraction. Morales and Roggen \cite{cnnLSTMHAR} proposed a model based on convolutional neural networks (CNNs) and long short-term memory (LSTM) components. In the work, CNN is employed to capture local temporal relationships, while LSTM's memory states help in the learning of large time scale dependencies. In \cite{DBNHAR}, a hybrid method was developed in which the sequence of human behaviors was modeled using a deep belief network as the emission matrix of a Hidden Markov Model. \add{In \cite{chen2021modeling}, a deep network architecture using stage distillation with a teacher and a student network are used to allow the model to extract more and more useful features from the raw data.}
These approaches are capable of extracting features from data automatically and without the need for domain knowledge. However, they continue to need explicit labels to supervise the model's training.

\add{In addition to these pure hand-engineered and purely learned feature representations, recent work has also combined the two techniques. For example, \cite{QIN202080} utilized engineered wavelet features arranged in the form of an image but then applied this as the input to a convolutional network to automatically select and abstract these engineered base features to obtain a more powerful feature representation for the problem. This again gives more flexibility to the learning system but, like the other approaches still continues to rely on explicit labels for learning.}  

\add{In the work presented in this paper, the requirement for explicit labels is dramatically relaxed through the application of self-supervised an unsupervised learning components. To facilitate the architecture, a hybrid approach to input features is taken where statistical measures are derived from data segments and utilized as input to deep learning structures to select and adapt useful features for the activity recognition task.}

\add{For a more detailed discussion of the different aspects and methods in HAR, please refer to two recent survey papers in the area, \cite{MINHDANG2020107561} and \cite{WANG20193}.}

\subsection{ResNet Autoencoder and Siamese Networks}
We take ideas from \com{Can we include references for these three components here ?}Autoencoder \cite{Kramer1991NonlinearPC}, ResNet \cite{resnet} and siamese architectures \cite{signature1993} to construct our model that is capable of effectively capturing patterns in data and computing semantic similarity between pairs of data samples.

\subsubsection{Autoencoder}
Autoencoder is a highly effective and extensively used unsupervised deep learning architecture for feature learning. It is divided into two components: an encoder and a decoder. The encoder specifies the nonlinear transformation $E(\cdot)$ that is used to convert the input data sample $x$ to the representation $E(x)$. The decoder specifies another nonlinear transformation $D(\cdot)$ with the goal of decoding $E(x)$ and reconstructing the original input $x$.
\begin{align}
\tilde{x} = D(E(x))
\end{align}
Here, $x$ denotes the input data, $E(x)$ denotes the encoded representation from the encoder, and $\tilde{x}$ is the decoded reconstruction from the decoder. The learning target of a conventional autoencoder is to minimize the reconstruction loss:
\begin{align}
\Phi_{ae}(x) = \sum_{x \in X}  ||x - \tilde{x}||^2
\end{align}
where $X$ denotes the dataset. This loss function is used to ensure that the reconstruction $\tilde{x}$ is as close to the original input $x$ as possible. If a decent reconstruction $\tilde{x}$ can be decoded from the representation $E(x)$, it indicates that the representation $E(x)$ has retained the essential information from the input $x$, allowing the reconstruction $\tilde{x}$ to be very similar to the original input $x$. As a result, the representation $E(x)$ may be utilized for further tasks \del{like}\add{such} as classification and clustering. 

Yet, \add{often the} general reconstruction objective is insufficient. The target of \add{a} classic autoencoder is to develop a representation $E(x)$ that has sufficient information to reconstruct the input, which implies that an accurate reconstruction contains noise and all the nuances in the input data. \del{Nevertheless}\add{However}, not all of the information included in the learnt representation is relevant to the downstream task (e.g. noise as well as some task-irrelevant details might not only be unnecessary but could even be detrimental, especially in the context of subsequent classification and clustering tasks). To develop an effective representation for categorizing activities effectively, the representation should include just the information necessary to perform the task while mostly disregarding unnecessary details. Based on this observation, our approach incorporates additional task-oriented loss functions to guide the learning process of the autoencoder and improve the efficacy of the learnt representations in subsequent tasks.

\subsubsection{ResNet} 
\begin{figure}
\begin{center}
\includegraphics[width=0.5\textwidth]{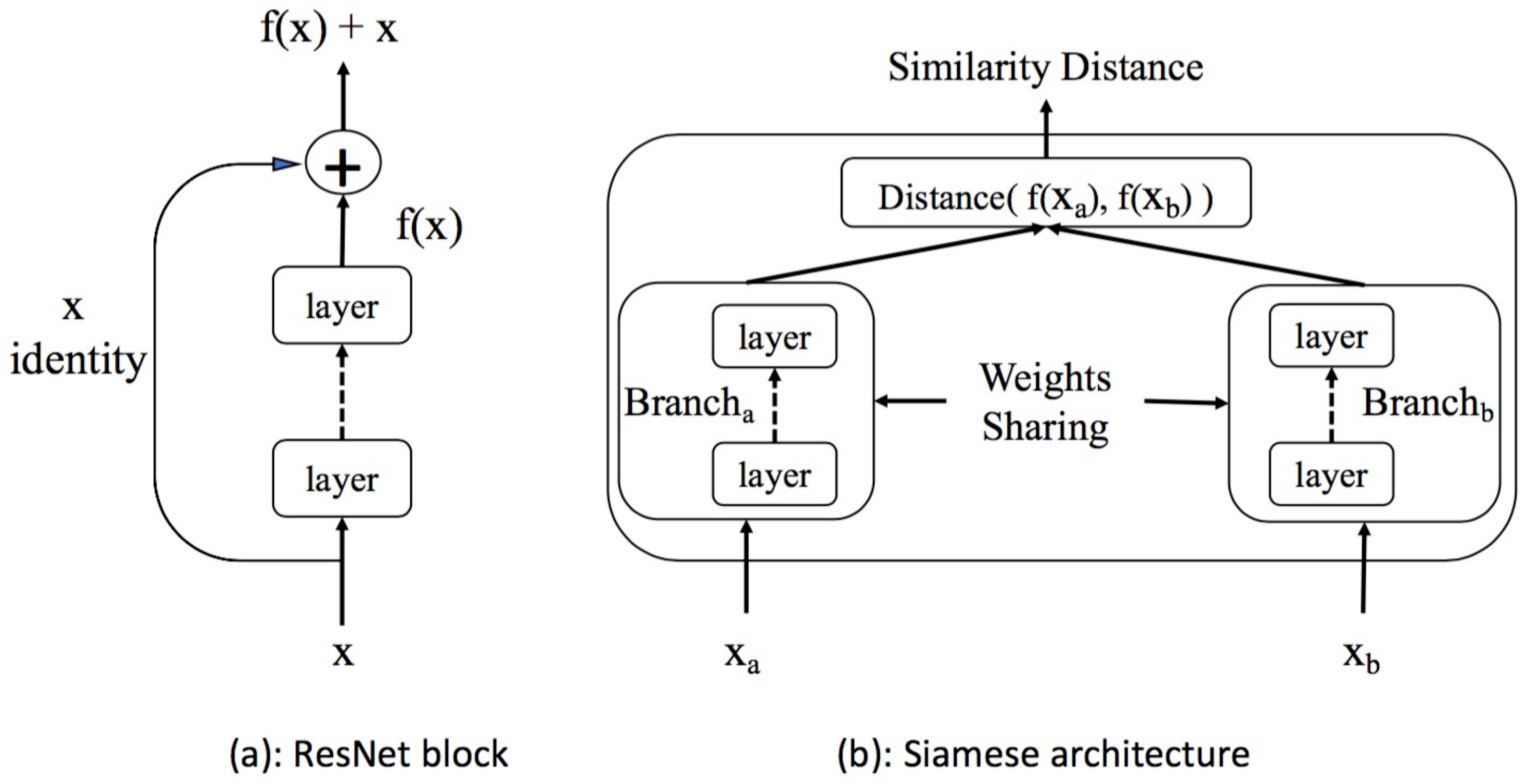}
\caption{ResNet and Siamese Architecture.} \label{resSiamese}
\end{center}
\end{figure}

ResNet allows the user, by utilizing skip connections or shortcuts to jump over some layers, \add{as shown conceptually fora single ResNet block in} Figure \ref{resSiamese}(a), to boost network capacity without suffering the expense of feature learning deterioration, when compared to standard deep neural networks. The output of a building block in ResNet is the sum of the result from the last non-linear layer and the input of the building block. However, in residual blocks, the shapes of the input tensor and the output tensor can be different. To address this problem, if the input tensor and output tensor have different shapes, a linear layer with the size of the output tensor will be used as the residual connection. Otherwise, an identity function will be used as the residual connection.

\subsubsection{Siamese Networks} 
Siamese Networks, as illustrated in Figure \ref{resSiamese}(b), are dual-branch neural networks with shared weights such that $Branch_{a} = Branch_{b}$. Given an input pair of human activity data samples $\{x_{a}, x_{b}\}$, the siamese network learns to project the input pair to the representation pair $\{E(x_{a}), E(x_{b})\}$. The distance between the representation pair is then used as the semantic similarity of the input pair.
Mueller and Thyagarajan \cite{sentenceSiamese} measured the semantic similarity between two sentences using siamese recurrent neural networks. In \cite{faceverification}, a siamese CNN is trained for face verification. Other applications include unsupervised acoustic model learning \cite{jointlearning,Synnaeve2016ATC,Kamper2015DeepCA}, image recognition \cite{eccvsiamese}, and object tracking \cite{Koch2015SiameseNN}.

\section{Proposed Approach}
Our proposed approach circumvents the lack of labeled data by leveraging properties pertaining to the nature of human activities. Specifically, our approach makes use of two kinds of relationships: temporal consistency of time series of human activity and feature consistency of human activity data in feature space. 

In this section, we first introduce the architecture of the proposed model. Then we show the definitions of various consistency criteria that are used in the model, and how the domain knowledge is employed in the model in the form of these consistencies.

\subsection{Architecture of the Proposed Model}
\begin{figure}
\begin{center}
\includegraphics[width=0.5\textwidth]{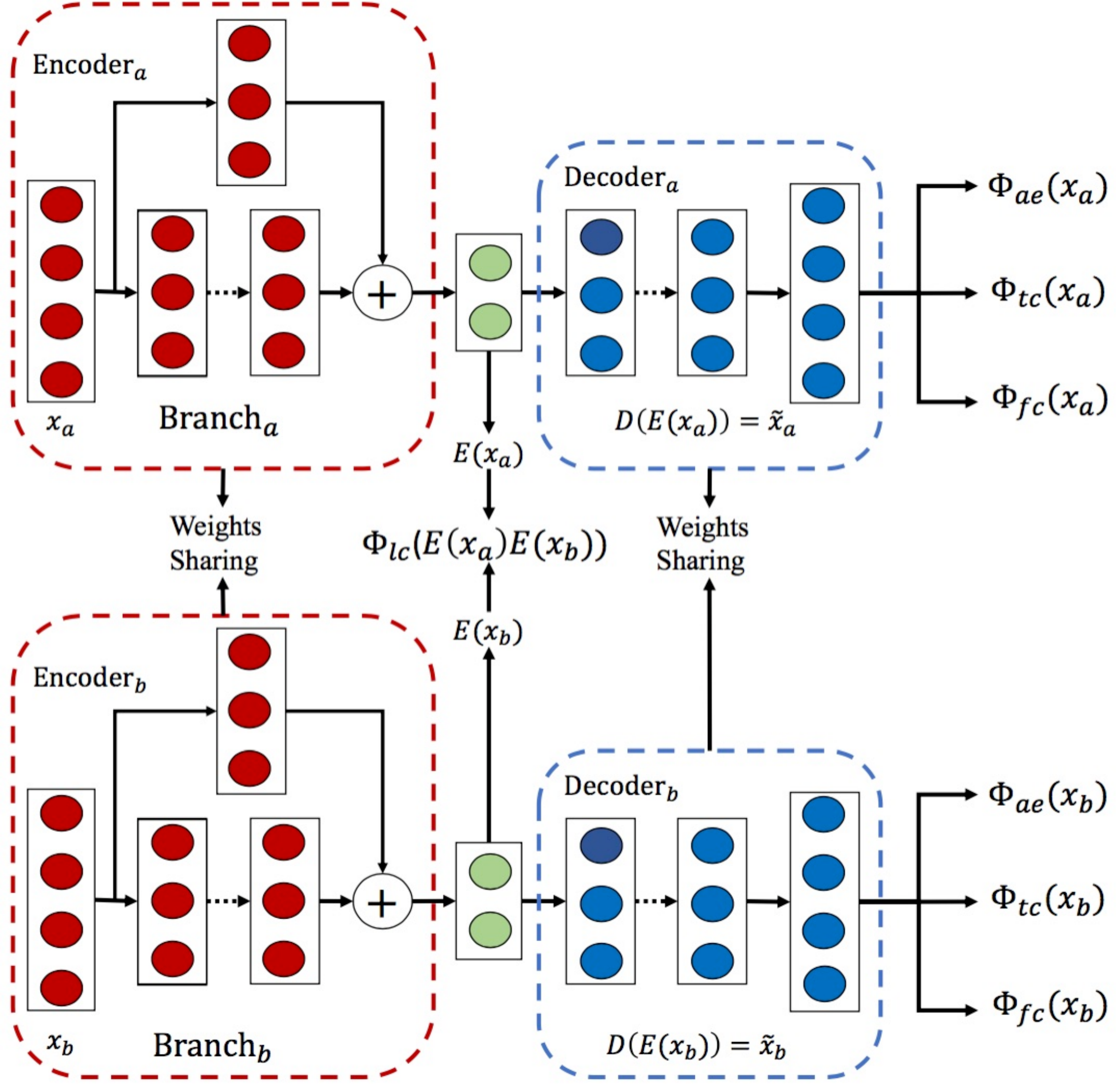}
\caption{The overall architecture of the proposed approach.} \label{chap6Architecture}
\end{center}
\end{figure}

As shown in Fig. \ref{chap6Architecture}, the foundation of the overall architecture of this approach is a combination of a ResNet autoencoder and siamese networks. 
The proposed approach utilizes self-supervised consistency criteria within the autoencoder components. Multiple learning targets, based on the nature of human activity, imposed on the autoencoder allow it to capture information that is relevant to the activity recognition task, and disregard irrelevant details in the data. In addition, the siamese network structure further enhances the recognition performance of the model with only limited amounts of weakly supervised data. Moreover, the residual connections, as mentioned above, increase the learning capacity of the model. These consistency criteria complement each other, obtaining a high accuracy model in a few-shot learning fashion.

\subsection{Consistency Definition}
The consistencies employed in the model are the common sense knowledge or domain knowledge, which can provide useful task-oriented information to the model, guide the model to learn task-relevant information, and ignore task-irrelevant information, hence improving the performance of the model.

\textbf{Temporal Consistency} describes the temporal continuity in human activity. Generally, human activities may be split intuitively into two components: a time variable component and a temporally fixed component. To be more precise, some dynamic features of a particular action may change over time. For instance, during walking, the body's posture changes with time: the left and right feet alternately advance. This sort of dynamic feature is also captured in sensor data and is referred to as the temporally changing component in this section.

On the other hand, regardless of how the bodily stance changes over time, the activity's semantic meaning stays constant. Specifically, the left and right feet may move forward alternately, but the action remains walking. This \del{section}\add{part} is referred to as the temporally stationary component.

Due to the nature of human activity, the temporal consistency loss pushes temporally near data samples to be similar to one another and ignores the temporal variable component difference. It is inspired by the goal of maintaining a reasonably stable semantic content, i.e. the sort of action in which we are engaged, across time. Though the data samples are temporally near, they may indicate the same sort of activity, even if they are quite far in the sensor data space in terms of Euclidean distance. The temporal consistency loss retains the sensor data's temporal continuity.

Formally, let $x_{i}^{t}$ denote the $i$-th data sample that occurs at time $t$ during the course of an activity. Let $X_{i}^{t}$ denote the set of $m$ data samples which are the nearest neighbors of $x_{i}^{t}$ based on the temporal closeness during the course of an activity, and $\tilde{x_{i}^{t}}$ the reconstruction of $x_{i}^{t}$. \add{Then the temporal consistency loss, $\Phi_{tc}(x_{i}^{t})$, is defined as:}
\begin{align}
\Phi_{tc}(x_{i}^{t}) = \frac{1}{m}\sum_{x_{j} \in X_{i}^{t}} ||x_{j} - \tilde{x_{i}^{t}}||^2
\end{align}
Temporal consistency attempts to structure data such that data samples that occur directly after each other are closer together in representation space, representing the intuition that activities are usually longer than a single data sample and thus that consecutive data samples are often still part of the same activity.

\textbf{Feature Consistency} was motivated by the fact that although various individuals conduct the same type of activity in a variety of ways, the variety of ways does not prevent other individuals from identifying the activity type. As a result, we think that the individual level characteristics of the activity data may not be required during the activity clustering stage, and that the characteristics that are consistently present across several data points may represent the activity's fundamental characteristics. This assumption underpins the Feature Consistency target function.

Previously published research \cite{Sheng2020UnsupervisedEL} \com{There should be a reference here. If it is your paper you can include "(reference withheld for double-blind review)".} has demonstrated that combining carefully designed handcrafted high level features to represent the primary characteristics of a temporally varying signal value with the k-Nearest Neighbor algorithm is an effective method for classifying sensor data from human activities. It is used in this strategy to define the local neighborhood of a data sample because \del{to}\add{of} its efficacy and simplicity. The Feature Consistency loss then attempts to retain the high-level feature traits that are typically present in a neighborhood.

The Feature Consistency loss compels the decoder to decode a data sample using its neighboring data samples' learnt representation. The reasoning behind this is because if data samples are spatially adjacent to one another in the constructed feature space, they may indicate the same sort of activity. Thus, the characteristics shared by several neighboring data samples should be considered to represent the key characteristics of that type of activity. If the features are not present in all surrounding data sets, they may indicate individual level characteristics, but not activity characteristics.

Formally, let $x_{i}^{f}$ denote the $i$-th data sample in feature space $f$, \del{and} $X_{i}^{f}$ the set of $n$ data samples which are the nearest neighbors of $x_{i}^{f}$ in the feature space $f$, and $\tilde{x_{i}^{f}}$ the reconstruction of $x_{i}^{f}$. \add{Then the feature consistency loss, $\Phi_{fc}(x_{i}^{f})$, is defined as:}
\begin{align}
\Phi_{fc}(x_{i}^{f}) = \frac{1}{n}\sum_{x_{k} \in X_{i}^{f}} ||x_{k} - \tilde{x_{i}^{f}}||^2
\end{align}
Feature consistency aims to keep data samples that have very similar features close together in the embedding space. This reflects the intuition that data segments that exhibit very similar sensor readings are more likely to belong to the same activity.

\textbf{Label Consistency}: With the help of siamese networks and a very small amount of weakly-supervised data, a group of pairwise constraints can be defined on the input data. This group of pairwise constraints can drive the proposed model to map the data into a representation space, in which the clusters satisfy the given constraints. 

More formally, given an input pair of human activity data sequences $\{x_{a}, x_{b}\}$, the encoder of the siamese networks learns to encode the input pair to the representation pair $\{E(x_{a}), E(x_{b})\}$. Then, a distance function, which measures the similarity between the encoded representations, is calculated. For the case of Euclidean distance, the label consistency loss can then be defined as:
\begin{align}
\Phi_{lc}(x_{a}, x_{b}) = ||E(x_{a}) - E(x_{b})||^2
\end{align}
The similarity distance $Dist(x_{a}, x_{b})$ indicates if the input pair represents the same type of activity. A large distance indicates two different types of activities, while a small distance indicates the same types of activity.

The aforementioned three types of consistency criteria imposed on the proposed model enable the learning to capture the information that is useful to the activity recognition task.

\subsection{Joint Loss Function}
The training process of the model is divided into two stages. In the first stage the model is trained with the following joint loss function:
\begin{align}
\begin{split}
&min\sum_{i=1}^{S}(1-\alpha-\beta) \cdot \Phi_{ae}(x_{i}) + \alpha \cdot \Phi_{tc}(x_{i}^{t}) + \beta \cdot \Phi_{fc}(x_{i}^{f})
\end{split}
\end{align}
where $i$ is the index of the sample, $S$ is the size of the dataset, $\alpha$ and $\beta$ are the parameters to balance the contribution of $\Phi_{ae}$, $\Phi_{tc}$, and $\Phi_{fc}$. While $\Phi_{tc}$ and $\Phi_{fc}$ retain task-relevant features and disregard the unnecessary task-irrelevant features, the $\Phi_{ae}$ component is also necessary in the learning process because without the reconstruction loss $\Phi_{ae}$, the risk of learning trivial solutions or worse representations will be increased \cite{taxonomy}. This represents an\del{d} initial self-supervised training phase in which the embedding space is trained to reflect the basic structure of the data.

In the second training stage, the label consistency loss $\Phi_{lc}$ is added to further improve the performance of the model. Since in the first stage, most clusters are already formed, the second stage imposes a small amount of pairwise constraints on the data using the limited amount of available weakly supervised data to readjust the already formed clusters to adapt to the available information about the intended classes. The joint loss function is defined as follow:
\begin{align}
\begin{split}
&min\sum_{i=1}^{S} (1-\alpha-\beta-\gamma) \cdot (\Phi_{ae}(x_{a})+ \Phi_{ae}(x_{b})) + \\
&\alpha \cdot (\Phi_{tc}(x_{a}^{t}) + \Phi_{tc}(x_{b}^{t})) + \\
&\beta \cdot (\Phi_{fc}(x_{a}^{f}) + \Phi_{fc}(x_{b}^{f})) + \\
&\gamma \cdot \Phi_{lc}(x_{a}, x_{b})
\end{split}
\end{align}
The label consistency supervision $\Phi_{lc}$ used in the model is limited, but it is directly related to the activity recognition task, so in the second stage, the label consistency will have a larger weight and dominate the learning process, and the other loss functions will have smaller weights, so that the model can learn directly task-related features. Additionally, the amount of label consistency supervision used in the model is very limited, therefore, the other two self-supervised and one unsupervised loss functions are still needed to work as regularization terms to prevent potential overfitting.

\subsection{Contrastive Loss Function}
Part of the objective of the model is to minimize the reconstruction error\del{,} \add{and to optimize} temporal and feature consistency using Euclidean distance,
while a contrastive loss function \cite{contrast} is used for \del{the} label consistency.

Given the input pair $\{x_a, x_b\}$, the model outputs a pair of activity representations, $\{E(x_a), E(x_b)\}$. The contrastive loss function used to train the model is also defined on the pairs. The similarity distance $Dist_s$ between the pair of data samples is calculated using the euclidean distance between their representation pair,
\begin{align}
& Dist_{s}(x_a, x_b) = ||E(x_a)-E(x_b)||_2
\end{align}
\del{Here}\add{In the following}, $Dist_{s}(x_a, x_b)$ is rewritten as $D$ to simplify the notation. Then, for each of the categorizations used in training, the loss function is defined as follows\del{,}\add{:}
\begin{align}
& L^i(x_{a}^{i}, x_{b}^{i}, y^i) = y^{i}L_{s}( x_{a}^{i}, x_{b}^{i})+(1-y^{i})L_{d}(x_{a}^{i}, x_{b}^{i}) \\
& L(x_a, x_b, y) = \sum_{i=1}^{N} L^i(x_{a}^{i}, x_{b}^{i}, y^i)
\end{align}
where $(x_{a}^{i}, x_{b}^{i}, y^i)$ is the $i$-th sample in the data set. $L_d$, $L_s$ are the loss terms for the negative pair\add{s} of data samples $(y=0)$ and the positive pair\add{s} of data samples $(y=1)$ respectively. The forms of $L_d$ and $L_s$ are given as,
\begin{align}
&L_d=\frac{1}{2}\{max(0, \delta-D)\}^2 \\
&L_s = \frac{1}{2}(D)^2
\end{align}
where $\delta$ is a hyperparameter for the margin. It states that \add{only} if the distance between two negative data samples is less than the margin $\delta$, they contribute to the loss function.

\subsection{Feature Extraction}
Following the description of the proposed model, this part discusses the features used in experiments. The feature extraction step converts the segmented raw sensor inputs to feature vectors.
 
Let $r_i$ be the $sample_i$ in the set of segmented raw sensor signals, $x_i$ represent the converted feature vector, and $C$ denote the feature extraction function. Then, the extraction of features is defined as follows,
\begin{align}
x_i = C(r_i)
\end{align}
The proposed model's input is $x_i$. Table \ref{chap6Feat} summarizes the statistical high level features that are used in the proposed model. The approach makes use of the mean, variance, standard deviation, and median, which are the most often utilized features in HAR studies. Additionally, several other features that have been demonstrated to be effective in prior research works \cite{transitionAware} are included here as well. For instance, the feature interquartile range ($iqr$)\del{, which} is based on Quartiles ($Q_1$, $Q_2$ and $Q_3$) that divide the time series signal into quarters. Using these, $iqr$ is the measure of variability between the upper and lower quartiles, $iqr = Q_3 - Q_1$. 

Each of these features is calculated independently for each axis. Because the data from several sensors is synchronized, it is possible to combine data from multiple sensors. During the training process, the proposed model accepts these derived features as input and learns to retain the task-oriented information included in the features while discarding the task-irrelevant information.

\begin{table}[t]
\centering
\caption{List of the used statistical features.}\label{chap6Feat}\smallskip
\scalebox{0.9}{%
\begin{tabular}{|c|c|}
\hline
\textbf{Feature extraction function} &  \textbf{Description} \\
\hline
$mean(r_i) = \frac{1}{N}\sum_{j=1}^{N} r_{i_j} $ & Mean \\
%\hline
$var(r_i) = \frac{1}{N}\sum_{j=1}^{N} (r_{i_j} - mean(r_i))^2 $ & Variance \\
%\hline
$std(r_i) = \sqrt{var(r_i)} $  & Standard deviation \\
%\hline
$median(r_i)$ & Median values\\
%\hline
$max(r_i)$ & Largest values in array \\
%\hline
$min(r_i)$ & Smallest value in array \\
%\hline
$iqr(r_i) = Q_{3}(r_i) - Q_{1}(r_i) $ & Interquartile range \\
\hline
\end{tabular}}
\end{table}

\subsection{Cluster Construction}
The network is trained using the standard backpropagation approach with stochastic gradient descent. Each weight is first set to a small value. The initial learning rate is $lr = 0.05$, and after every 10000 steps, an exponential decay function with a decay rate of 0.95 is applied to the learning rate. The network is evaluated using validation data at the end of each epoch. The training procedure is complete when the validation error has not decreased for a predetermined number of epochs.

\begin{figure}
\begin{center}
\includegraphics[width=0.45\textwidth]{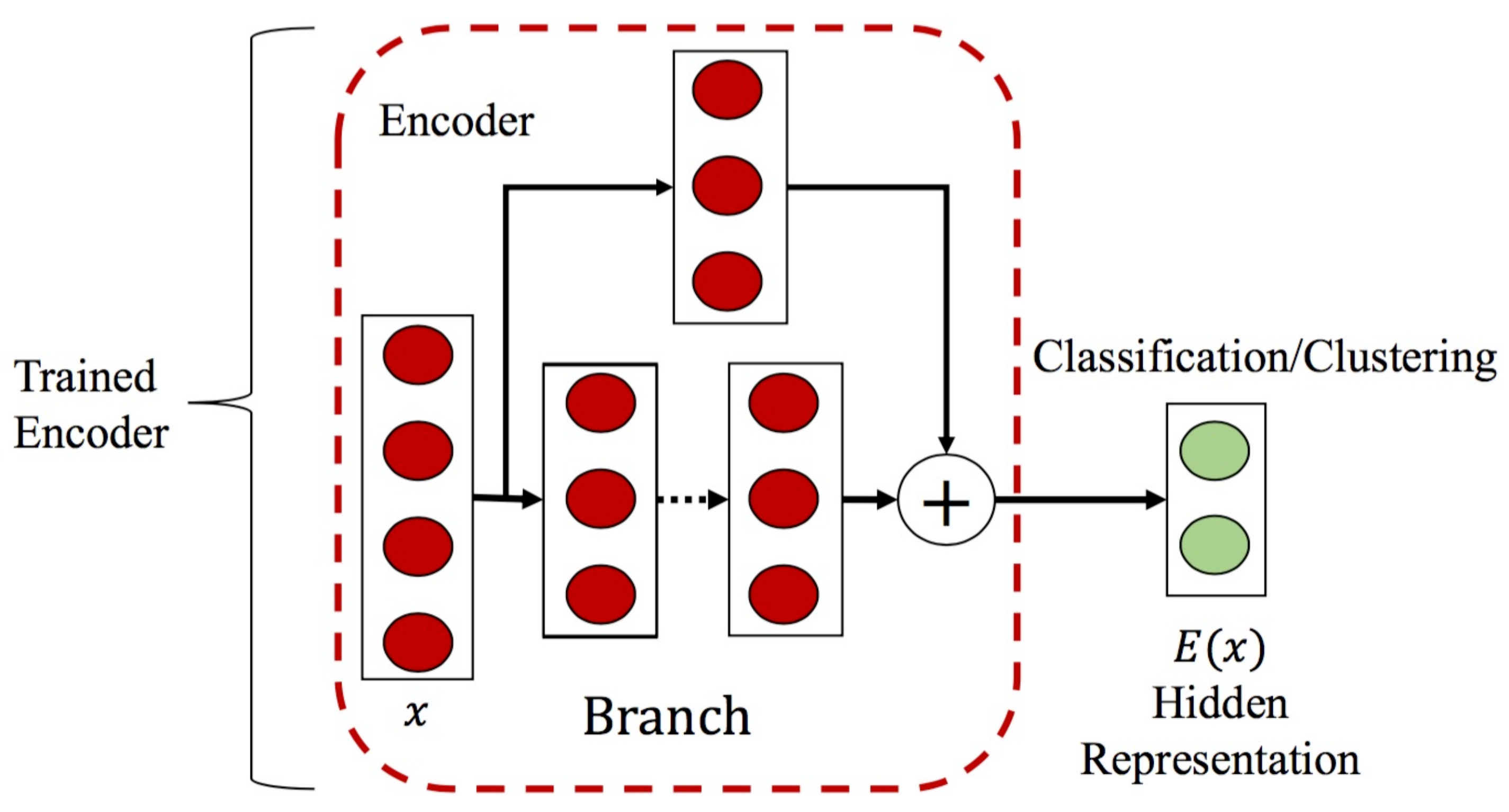}
\caption{Trained Encoder for Classification/Clustering.} \label{encoder}
\end{center}
\end{figure}

After two stages of training to build the hidden representation within the siamese and ResNet autoencoder architecture, the trained model (the encoder of the model, as shown in Figure \ref{encoder}) can project the input data sample $x$ into a clustering-friendly embedding space. The learnt representations are evaluated in the subsequent clustering task. The experiments use $k$-means (KM), arguably the most widely used clustering algorithm. The number of clusters in each dataset is selected to equal to the true number of classes.

\section{Evaluations and Experiments}
Three publicly available benchmark datasets containing wearable sensor data from various human activities are utilized to evaluate the proposed model.
\subsection{Datasets}
The three HAR datasets include: PAMAP2 \cite{pamap2}, REALDISP \cite{realdisp}, and SBHAR \cite{sbhar}. The sliding window technique is used to segment all sensor data sequences.

\textbf{PAMAP2} is a dataset collected from nine volunteers who completed 12 activities while wearing three inertial measurement devices (IMU) on their wrist, chest, and ankle. The dataset includes data on sporting activities (rope jumping, nordic walking, and so on) as well as home activities (vacuum cleaning, ironing etc.). Heart rate, accelerometer, gyroscope, magnetometer, and temperature data are collected during the experiments. In accordance with previous studies on this dataset, the data is segmented using a sliding window of 5.12 seconds with a one-second step size.

\textbf{SBHAR} collects data from 30 subjects executing 6 basic activities such as walking and laying, as well as 6 postural transitions such as stand-to-sit and sit-to-lie. In our experiment, we treat all postural transitions as \del{a} one general transition. The data was collected by strapping a smartphone to the participant's waist and recording accelerometer and gyroscope data at a sampling rate of 50Hz using the inertial sensors on the smartphone. As in the previous study \cite{transitionAware}, we employ a sliding window with a step size of 1.28 seconds and a duration of 2.56 seconds.

\textbf{REALDISP} is collected from 17 individuals performing 33 activities with the use of 9 sensors attached to their arms, legs, and back. Each sensor measures acceleration, gyroscope rotation, and the direction of the magnetic field. This dataset provides data on fitness workouts, as well as data from the warm-up and cool-down periods. The sliding window utilized in this case is two seconds in length without overlap.

To evaluate the performance of the proposed model, \add{the original fully labeled data was modified by removing labels and randomly selecting} \del{it was applied with} 1\%, 5\%, and 10\% weakly supervised data\add{. The performance of the proposed approach with this information-reduced data} \del{ and its performance} was compared to fully supervised methods \add{that had access to the complete original dataset} as well as to \mdel{\add{an}}\del{the} unsupervised \madd{and purely self-supervised} approaches. \mdel{In addition,}\madd{The comparison of purely self-supervised technique here also provides some ablation results for the proposed approach as it is effectively} \mdel{we also demostrated the experiment results from} the self-supervised approach with only temporal and feature consistencies \cite{Sheng2020UnsupervisedEL}. In all experiments, the total dataset (i.e. unlabeled and labeled) comprised all data segments of the datasets, giving the supervised comparison approaches access to all the labels while our model has access only to similarity information for a very small subset. To apply the weakly self-supervised method, the set of data segments \mdel{were}\madd{was} split according to the above-mentioned percentage and labels were removed from the samples in the data sets. The small percentage of labeled data was then converted into weakly supervised data pairs which only contain information whether they belonged to the same or a different class.

\subsection{Results and Analysis}
\begin{figure*}
    \centering
    \begin{subfigure}[b]{0.33\linewidth}        %% or \columnwidth
        \includegraphics[width=\linewidth]{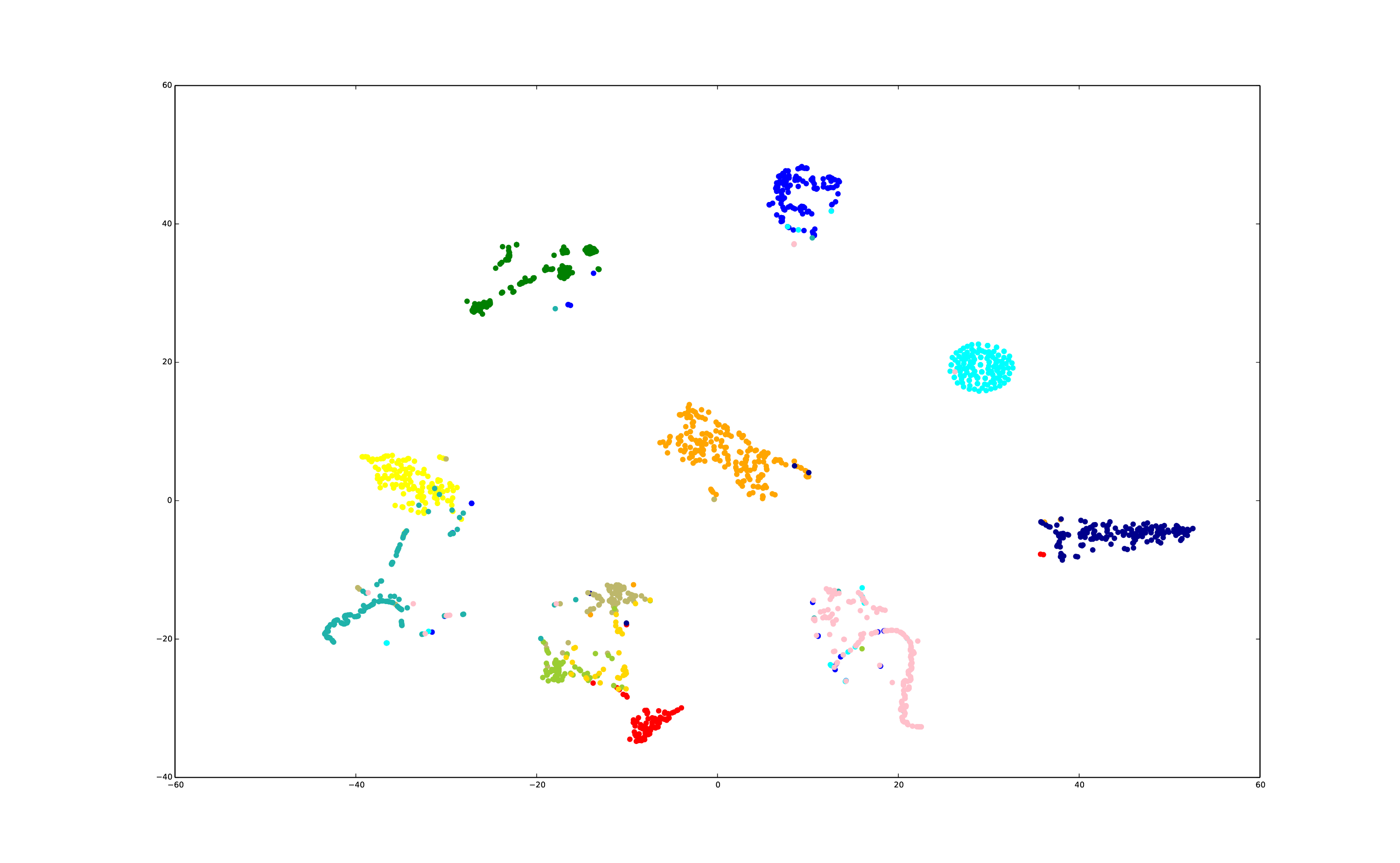}
        \caption{PAMAP2 with 1\% of labels}
        \label{PAMAP1percent}
    \end{subfigure}
    \begin{subfigure}[b]{0.33\linewidth}        %% or \columnwidth
        \includegraphics[width=\linewidth]{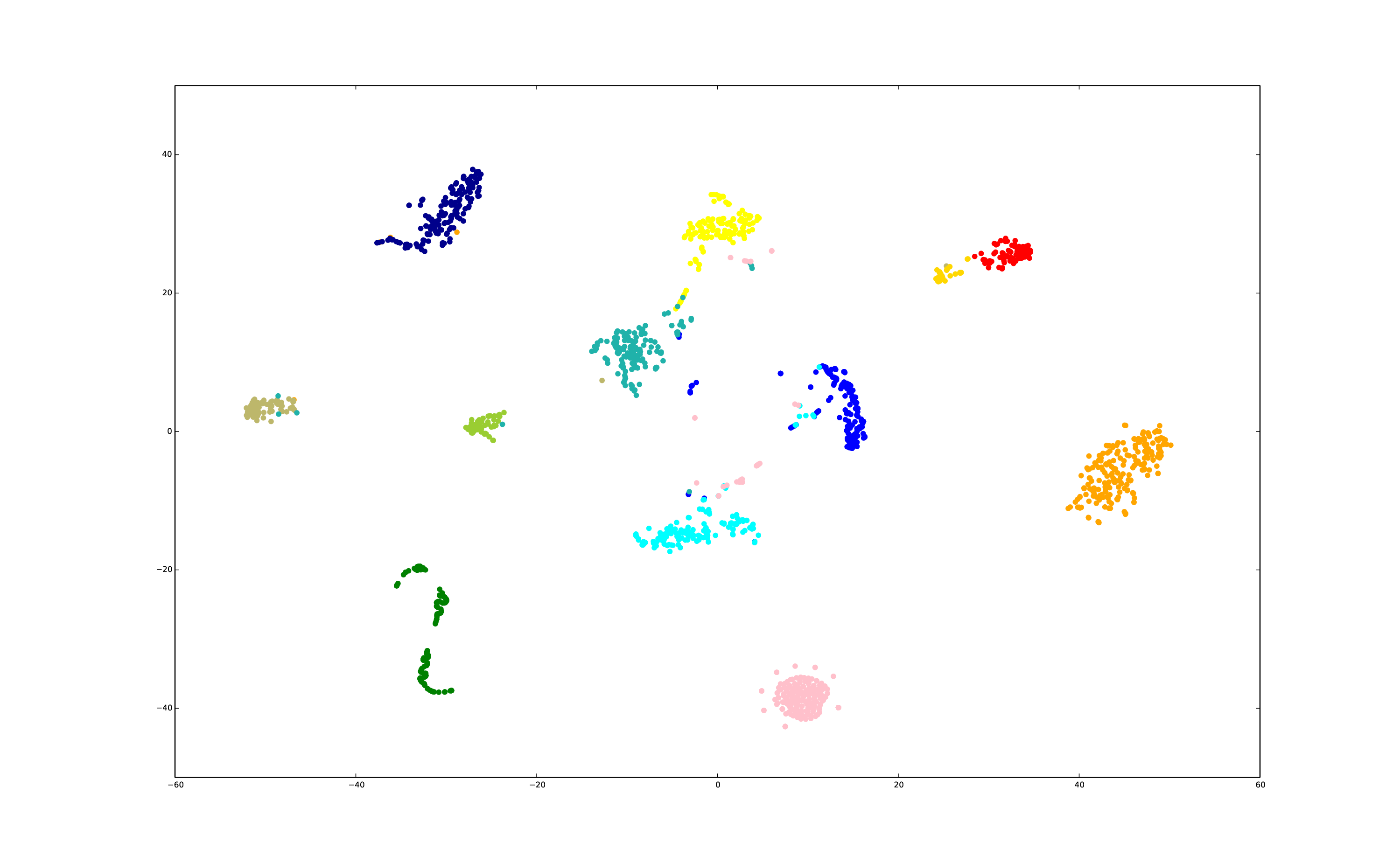}
        \caption{PAMAP2 with 5\% of labels}
        \label{PAMAP5percent}
    \end{subfigure}
     \begin{subfigure}[b]{0.33\linewidth}        %% or \columnwidth
        \includegraphics[width=\linewidth]{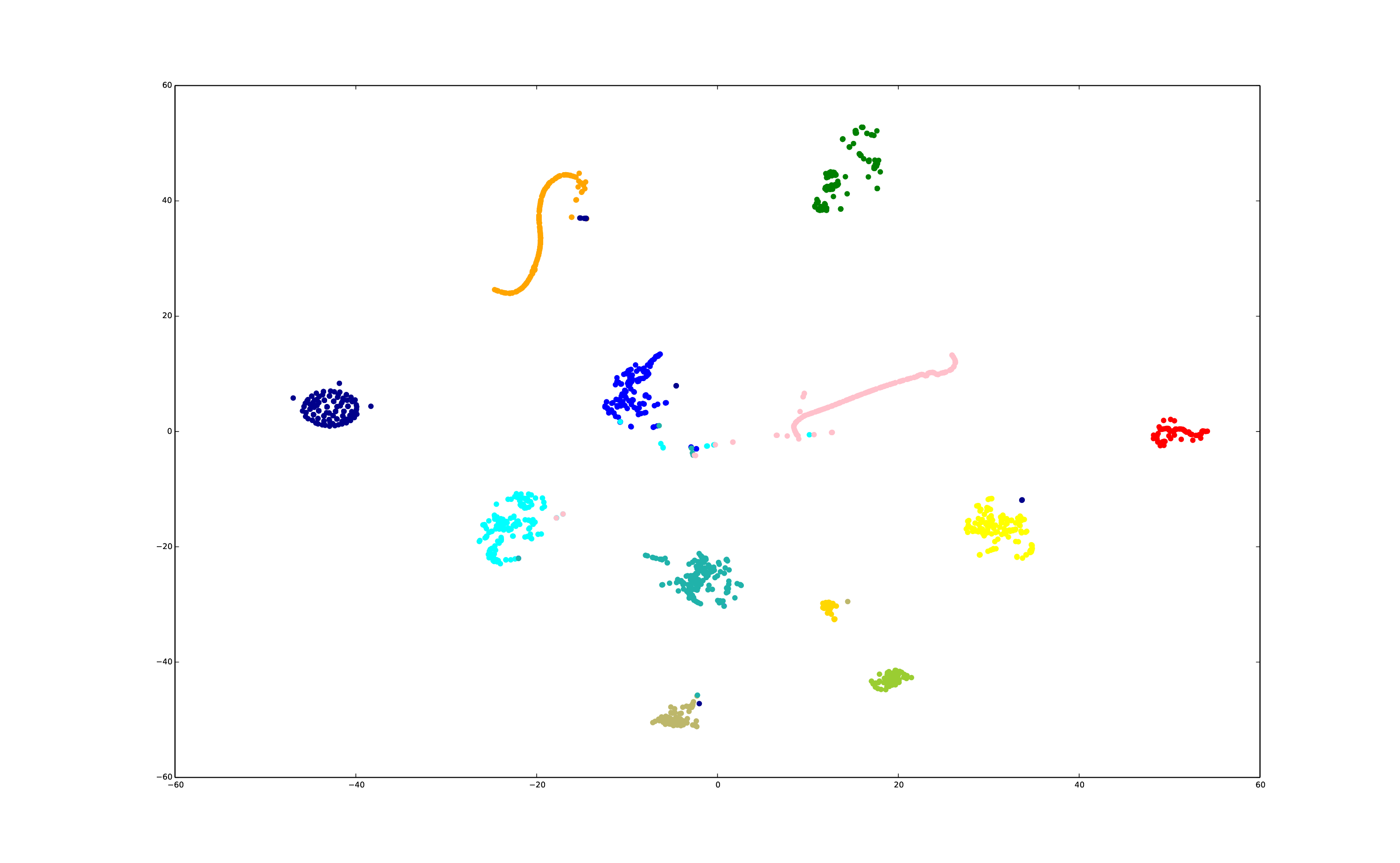}
        \caption{PAMAP2 with 10\% of labels}
        \label{PAMAP10percent}
    \end{subfigure}
    \caption{Representation space visualizations on the PAMAP2 dataset.}\label{PAMAP}
\end{figure*}

\begin{figure*}
    \centering
    \begin{subfigure}[b]{0.33\linewidth}        %% or \columnwidth
        \includegraphics[width=\linewidth]{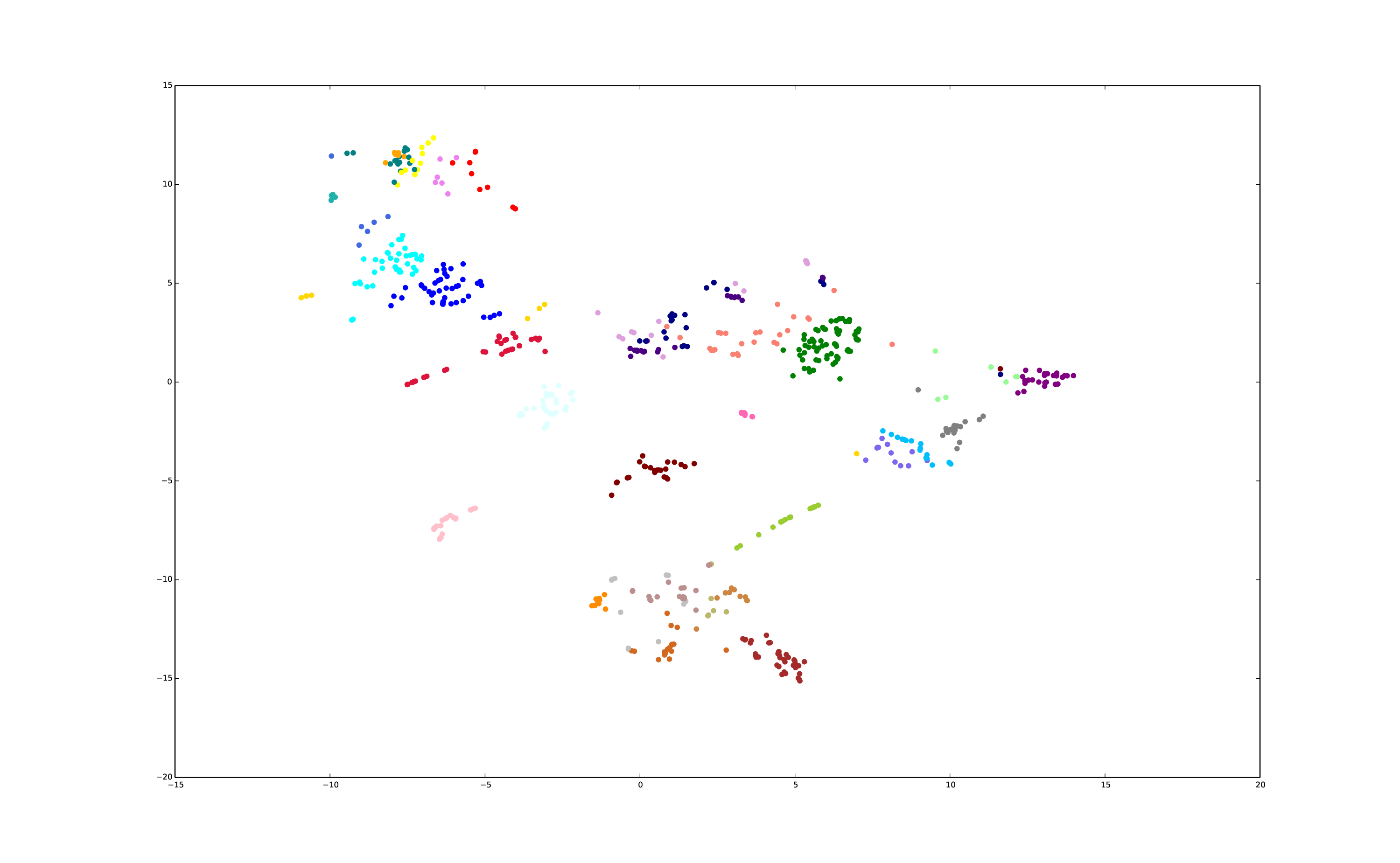}
        \caption{REALDISP with 1\% of labels}
        \label{REALDISP1percent}
    \end{subfigure}
    \begin{subfigure}[b]{0.33\linewidth}        %% or \columnwidth
        \includegraphics[width=\linewidth]{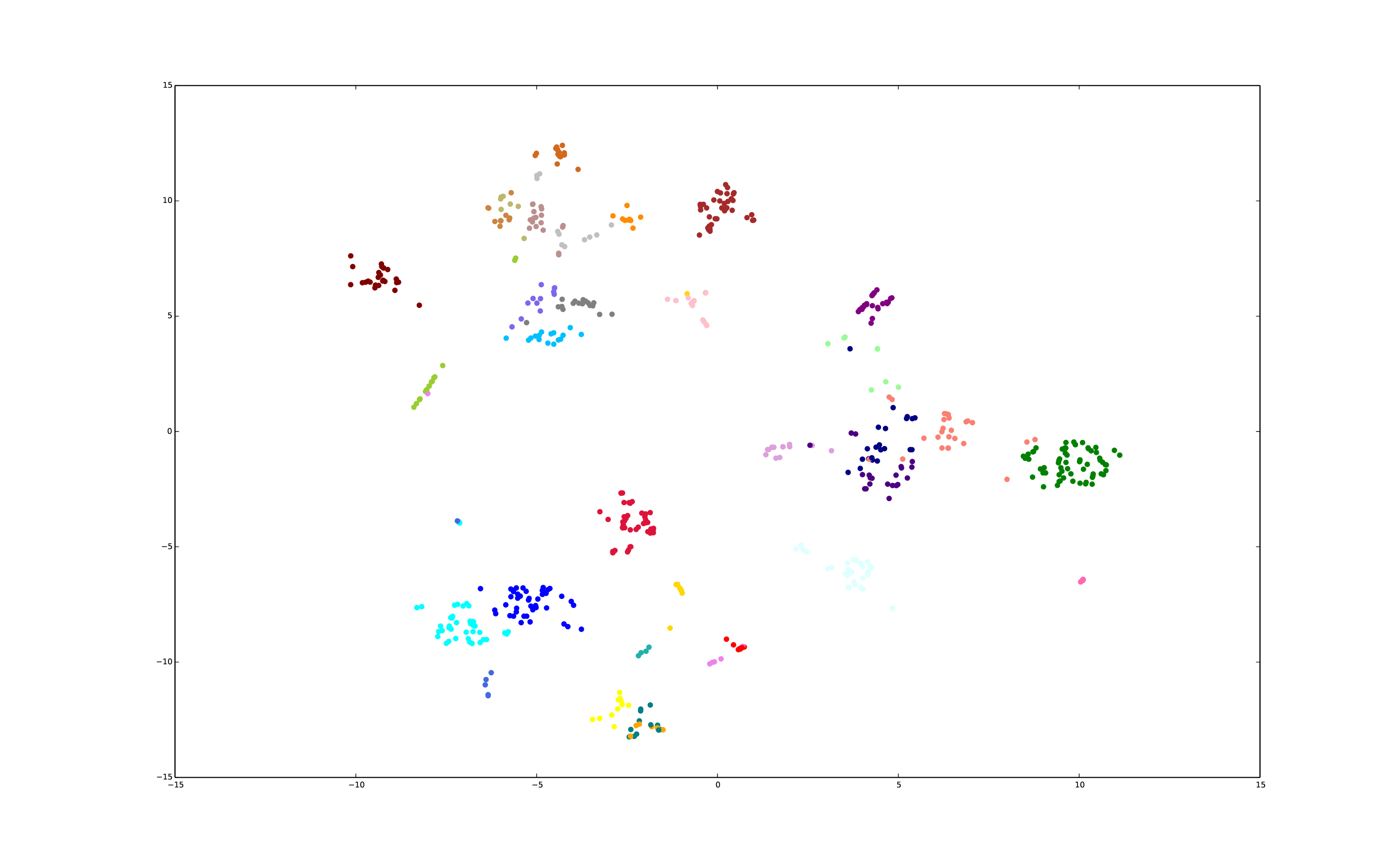}
        \caption{REALDISP with 5\% of labels}
        \label{REALDISP5percent}
    \end{subfigure}
     \begin{subfigure}[b]{0.33\linewidth}        %% or \columnwidth
        \includegraphics[width=\linewidth]{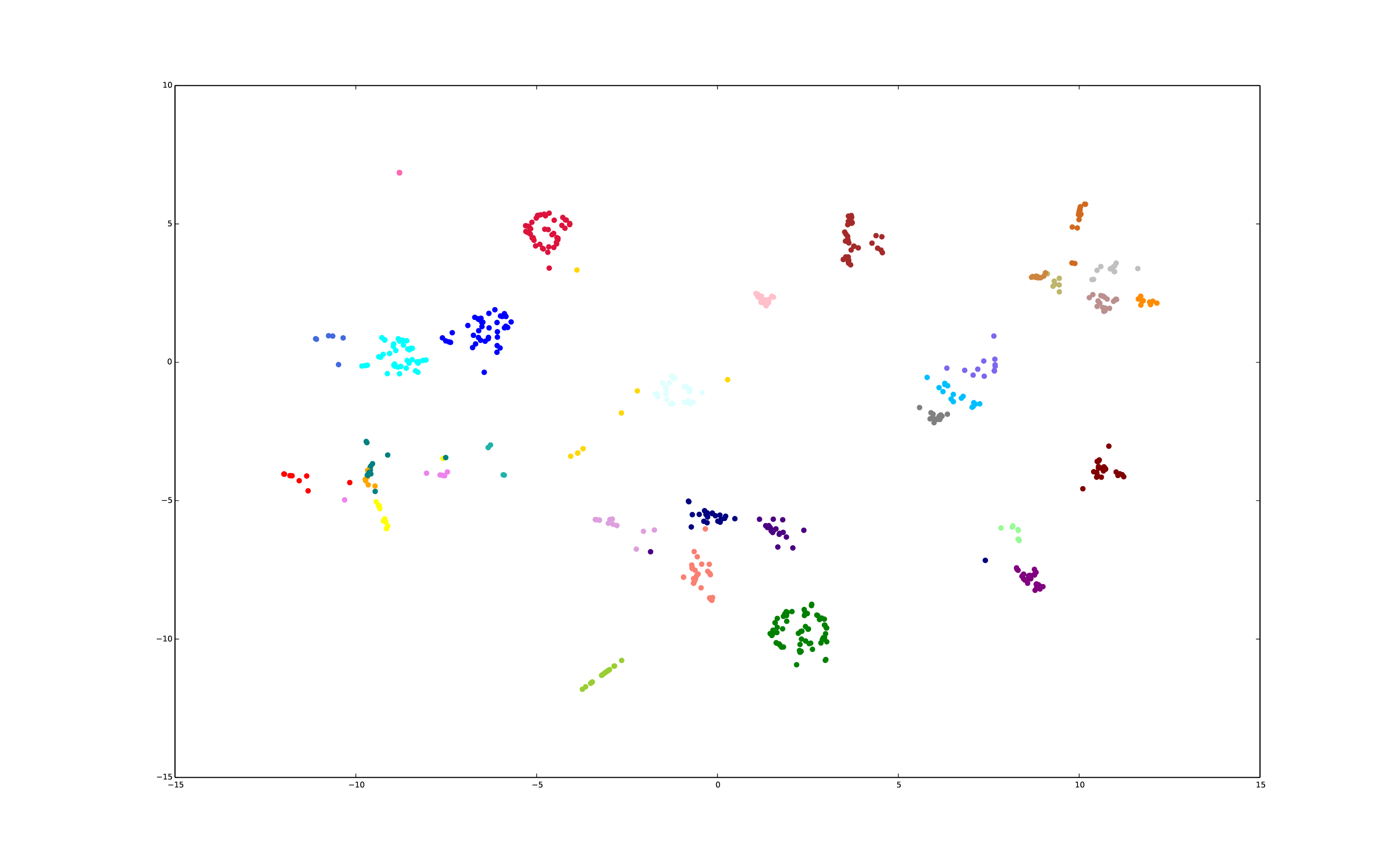}
        \caption{REALDISP with 10\% of labels}
        \label{REALDISP10percent}
    \end{subfigure}
    \caption{Representation space visualizations on the REALDISP dataset.}\label{REALDISP}
\end{figure*}

\begin{figure*}
    \centering
    \begin{subfigure}[b]{0.33\linewidth}        %% or \columnwidth
        \includegraphics[width=\linewidth]{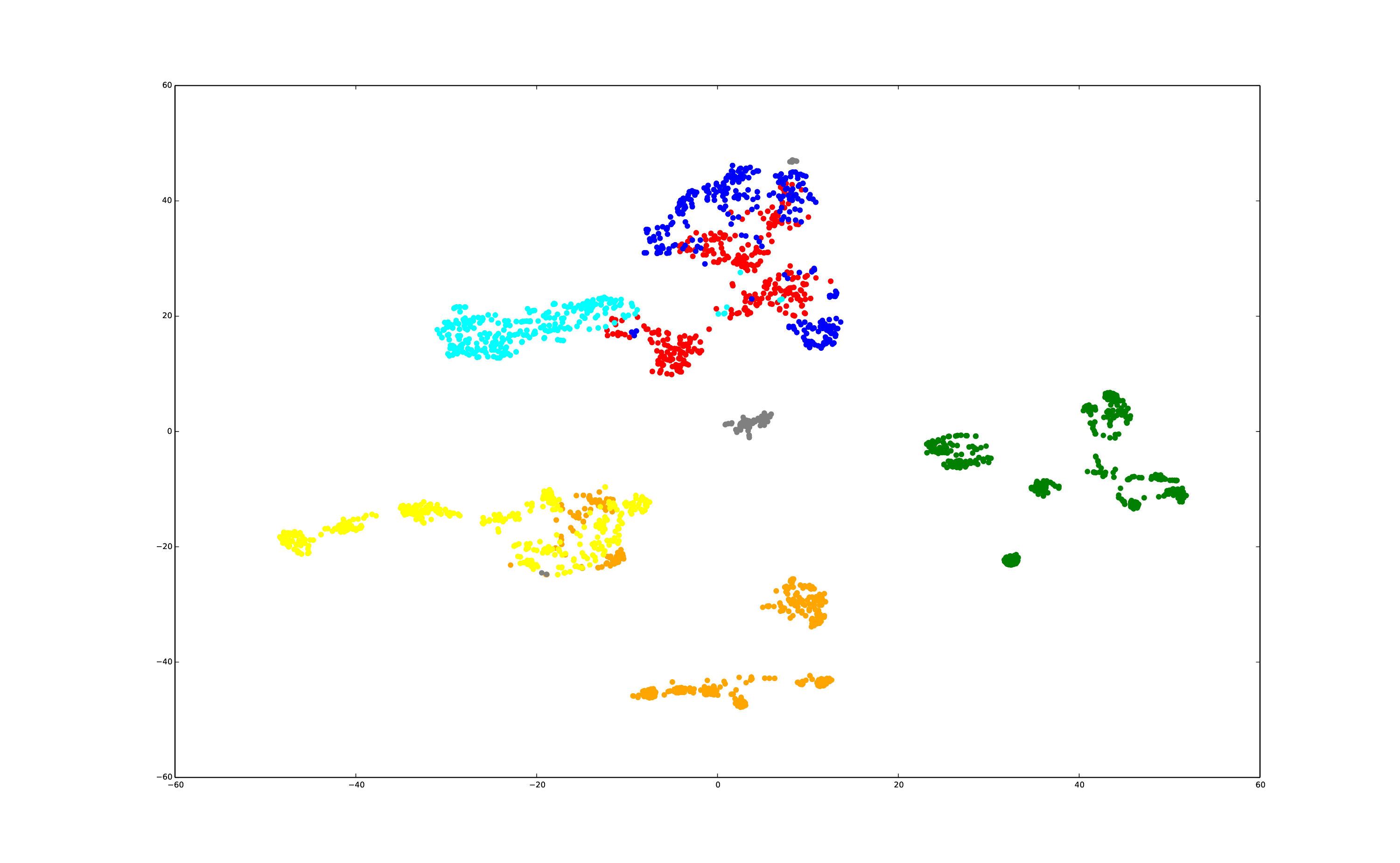}
        \caption{SBHAR with 1\% of labels}
        \label{SBHAR1percent}
    \end{subfigure}
    \begin{subfigure}[b]{0.33\linewidth}        %% or \columnwidth
        \includegraphics[width=\linewidth]{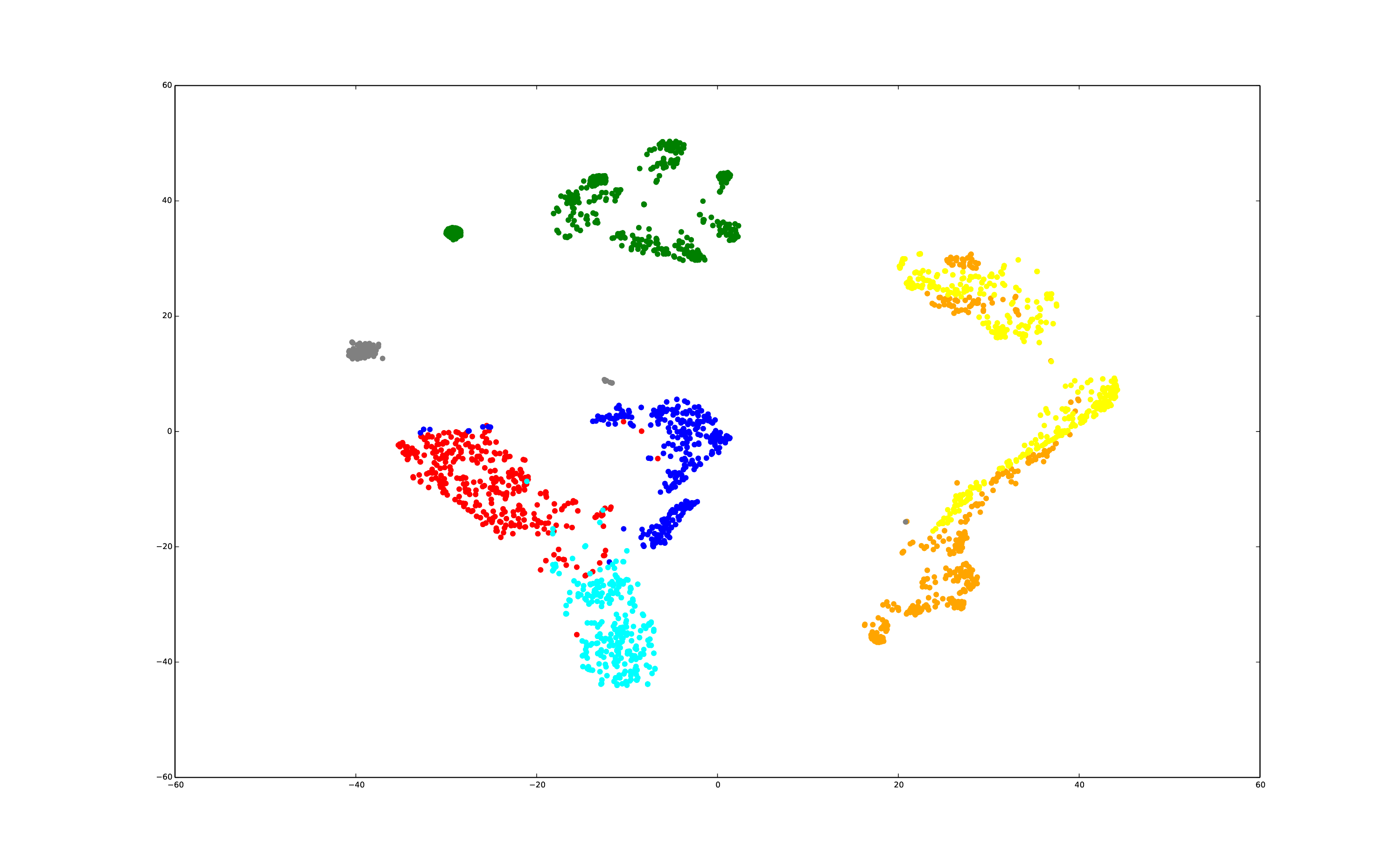}
        \caption{SBHAR with 5\% of labels}
        \label{SBHAR5percent}
    \end{subfigure}
     \begin{subfigure}[b]{0.33\linewidth}        %% or \columnwidth
        \includegraphics[width=\linewidth]{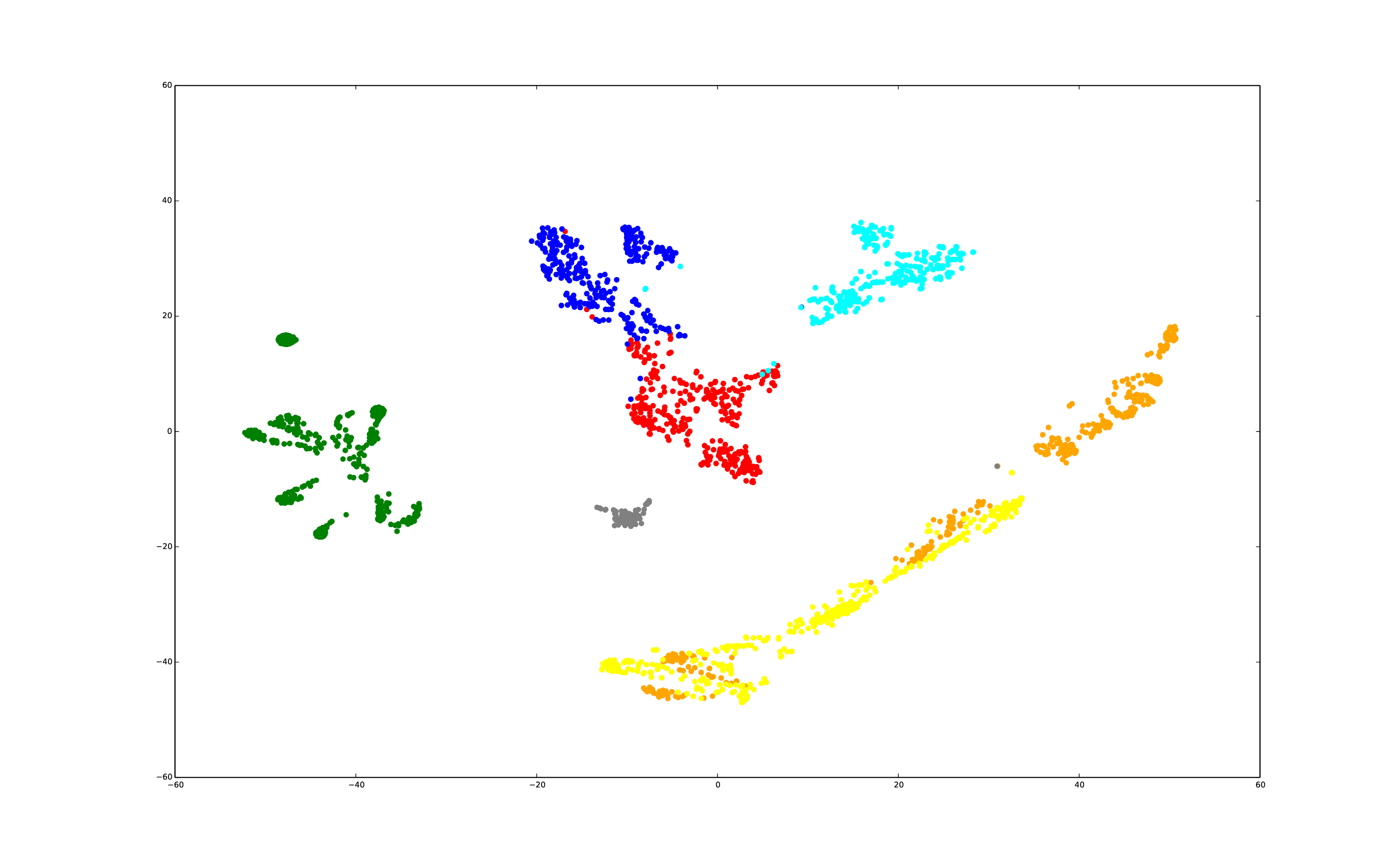}
        \caption{SBHAR with 10\% of labels}
        \label{SBHAR10percent}
    \end{subfigure}
    \caption{Representation space visualizations on the SBHAR dataset.}\label{sbhar}
\end{figure*}

To gain a deeper understanding of the model's learnt representations, we visualize the distribution of the activity vectors in the representation space. We used t-sne \cite{tsne} to map the representation vectors to two dimensions. The visualization results \add{for the PAMAP2, REALDISP, and SBHAR datasets} are shown in Figures \ref{PAMAP}, \ref{REALDISP}, and \ref{sbhar}, \add{respectively. The} various colors in the figures indicate distinct activity labels in the datasets. The experimental results \add{for the datasets} are summarized in Table\add{s} \ref{semi-pamap2}, \ref{semi-realdisp}, and \ref{semi-sbhar}. 

From the \del{the} figures and tables, we can reach the conclusions that the proposed weakly self-supervised approach achieved significant improvement over the purely unsupervised method with the same number of clusters even with very limited data and improved performance with the increase of available weakly supervised data. Moreover, performance was reasonable good even with only 1\% of the labeled training data. With 10\% of the labeled data, the proposed approach can achieve competitive performance compared with even the purely supervised methods, despite it having access to only a fraction of the labeled data.

\begin{table}
\centering
\caption{Results on PAMAP2}\label{semi-pamap2}
\vspace*{-0.15in}
\scalebox{0.85}{%
\begin{tabular}{|c|c|}
\hline
\textbf{Methods} &  \textbf{ACC}\\
\hline
MTLS \cite{Sheng2020WeaklySM} & 0.9893\\
Probability SVM with Filter \cite{transitionAware} &  0.9433 \\
Boosted C 4.5 \cite{pamap2} & 0.9969\\
%LSTM-F \cite{dnnHARbenchmark} & 0.9290 \\
CNN \cite{dnnHARbenchmark} & 0.9370 \\
UEL Autoencoder \cite{Sheng2020UnsupervisedEL} & 0.8543\\
Vanilla Autoencoder  & 0.7706 \\
% Unsupervised Learner in chapter \ref{chap4flairs20} + KM (TN)& 0.8543\\
% Unsupervised Learner in chapter \ref{chap4flairs20} + KM (TN+2)& 0.9211\\
Proposed approach with 1\% labels & 0.9128\\
Proposed approach with 5\% labels & 0.9755\\
Proposed approach with 10\% labels & 0.9904\\
\hline
\end{tabular}}
\end{table}

\begin{table}
\centering
\caption{Results on REALDISP}\label{semi-realdisp}
\vspace*{-0.15in}
\scalebox{0.85}{%
\begin{tabular}{|c|c|}
\hline
\textbf{Methods} &  \textbf{ACC}\\
\hline
Probability SVM with Filter \cite{transitionAware} & 0.9952 \\
Adaboost Ensemble Classifier \cite{realdispCompare} & 0.9998\\
CNN \cite{realdispCompare2} & 0.9280 \\
SMLDist \cite{chen2021modeling} & 0.9400 \\
UEL Autoencoder \cite{Sheng2020UnsupervisedEL} & 0.6812\\
Vanilla Autoencoder  & 0.6401 \\
% Unsupervised Learner in chapter \ref{chap4flairs20} + KM (TN)& 0.8543\\
% Unsupervised Learner in chapter \ref{chap4flairs20} + KM (TN+2)& 0.9211\\
Proposed approach with 1\% labels & 0.7515\\
Proposed approach with 5\% labels & 0.8780\\
Proposed approach with 10\% labels & 0.9223\\
\hline
\end{tabular}}
\end{table}

\begin{table}
\centering
\caption{Results on SBHAR}\label{semi-sbhar}
\vspace*{-0.15in}
\scalebox{0.85}{%
\begin{tabular}{|c|c|}
\hline
\textbf{Methods} &  \textbf{ACC}\\
\hline
MTLS \cite{Sheng2020WeaklySM} & 0.9885\\
Probability SVM with Filter\cite{transitionAware} & 0.9678 \\
tFFT + CNN \cite{Ronao2016HumanAR} & 0.9575\\
STLS \cite{Sheng2019SiameseNF} & 0.9268\\
UEL Autoencoder \cite{Sheng2020UnsupervisedEL} & 0.7401\\
Vanilla Autoencoder  & 0.6369 \\
% Unsupervised Learner in chapter \ref{chap4flairs20} + KM (TN)& 0.8543\\
% Unsupervised Learner in chapter \ref{chap4flairs20} + KM (TN+2)& 0.9211\\
Proposed approach with 1\% labels & 0.8015\\
Proposed approach with 5\% labels & 0.8853\\
Proposed approach with 10\% labels & 0.9352\\
\hline
\end{tabular}}
\end{table}

\section{Conclusions}
In this work, we have presented a weakly self-supervised embedding learning approach\del{,} which is based on a ResNet autoencoder \del{architecture} and a siamese architecture. The proposed approach uses t\add{wo generic}\del{he} properties of human activities: temporal consistency and feature consistency\add{,} to project the activity data into \add{an}\del{the} embedding space, \add{and} then imposes a small amount of pairwise constraints on the data to fine-tune the model in a weakly supervised fashion. We have demonstrated the effectiveness of the approach by applying it to three widely used HAR benchmark datasets \add{where for our approach labels were removed and only a very small percentage of data pairs were used as weakly supervised data by maintaining whether they were form the same activity class or not}. The results of the experiments show that with only ten percent of the labeled data \add{used for fine tuning through weak supervision}, the proposed approach can achieve competitive results, which significantly reduced the supervision needed to train the recognition model \add{and can thus dramatically reduce the overhead of obtaining useable datasets and extend the applicability of HAR to a broader range of ubiquitous sensor settings}.

\bigskip
\bibliography{aaai2020}

\end{document}